\documentstyle[prl,aps,multicol,epsf]{revtex}

\begin{document}

\draft
\tighten

\title{ Charge- versus spin driven stripe order: the role of transversal
spin fluctuations.}

\author{C. N. A. van Duin and J. Zaanen}
\address{Institute Lorentz for Theoretical Physics, Leiden University\\
P.O.B. 9506, 2300 RA Leiden, The Netherlands}
\date{\today ; E-mail:cvduin@lorentz.leidenuniv.nl; jan@lorentz.leidenuniv.nl}
\maketitle

\begin{abstract}
The separation of the charge- and spin ordering temperatures of the stripe
phase in cuprate superconductors has been used to argue that the striped
phase is charge driven. Scaling analysis of a non-linear sigma model shows that
the effect of spatial anisotropy on the transversal spin fluctuations is
much more drastic at finite temperatures than at zero temperature. These
results suggest that the spin fluctuations prohibit the spin system to 
condense at the charge ordering temperature, despite a possible
dominance of charge-spin coupling in the longitudinal channel.
\end{abstract}

\begin{multicols}{2}
\narrowtext
The observation of a novel type of
electronic order in cuprate superconductors and other doped
antiferromagnets has attracted considerable attention recently. 
In this stripe phase, the carriers are confined to
lines which are at the same time Ising domain walls in the N\'eel
background\cite{tranquadana}. 
Substantial evidence exists that dynamical stripe correlations
persist in the normal- and superconducting states of the 
cuprates\cite{tranquadaprl}. 

A further characterization of the fluctuation modes 
of the stripe phase is needed. In this regard, the finite temperature
evolution of the static stripe phase
might offer a clue. Both in cuprates\cite{tranquadana} 
and in nickelates\cite{tranquadani}, the
charge orders at a higher temperature than the spin, and both transitions
appear to be of second order. Zachar, Emery and Kivelson\cite{ZEM} 
argue on basis of a Landau free energy that
the stripe instability is {\em charge driven}: if the coupling between the
charge- and longitudinal spin mode would dominate, charge and spin would
order simultaneously in a first order transition. This is a mean-field
analysis, and fluctuations can change the picture drastically. For instance,
at length scales larger than the interstripe distance the spin system
remaining after the charge has ordered is just a quantum Heisenberg 
antiferromagnet in $2+1$ dimensions which cannot order at finite temperatures
according to the Mermin-Wagner theorem.  
Zachar {\em et al} argue that the orientational (`transversal') fluctuations
of the spin system can be neglected at the temperatures of interest,
because it appears that the spin system left behind after the
charge has ordered is not radically different from the antiferromagnet
in the half-filled cuprates, exhibiting a N\'eel temperature of order
300 K, an order of magnitude larger than that in the stripe phase.

An important constraint is that the $T=0$ staggered magnetization in
the stripe phase appears to be comparable to that at 
half filling\cite{tranquadaprl}. 
If the transversal fluctuations are responsible for the charge- and
spin transitions, it has to be demonstrated that the additional thermal
fluctuations due to the presence of stripes have a much greater effect on 
the N\'eel state than the $T=0$ quantum fluctuations. To investigate this,
we consider the simplest possible source of stripe induced spin disorder.
Following Castro Neto and Hone (CH)\cite{CH}
 we assume that the exchange coupling
between spins separated by a charge stripe is weaker than the interdomain
exchange, so that the collective spin fluctuations are described by
a spatially anisotropic $O(3)$ quantum non-linear sigma (AQNLS) model.
From our scaling analysis we find that a moderate anisotropy       
 (a factor of $\sim 4$ difference in spin wave
velocities) can explain a reduction of the N\'eel
temperature by an order of magnitude, while the $T=0$ staggered magnetization
is only reduced by a factor of two from its isotropic value. 
The reason can be inferred from the cross-over
diagram (Fig. 1). As function of increasing anisotropy, the $T=0$ transition
between the renormalized classical (RC) and quantum disordered (QD) states
scales to
smaller coupling constant, but the dimensionless temperature associated
with the crossover renormalized classical- to quantum critical scales
down much faster. Alternatively, we find that the behavior found by
Chakravarty, Nelson and Halperin(CHN)\cite{CHN}
 for the correlation length in the
RC regime of the isotropic model can be directly generalized to the
anisotropic case: the expression for the classical {\em anisotropic}
model remains valid when the bare stiffness is replaced by the renormalized
stiffness. It is suspected that this holds more generally. If so, the
strong disordering influence of temperature as compared to the quantum 
fluctuation might be generic: whatever the disordering influence of the
stripes is, it exerts it in an effectively three dimensional classical system 
at zero temperature and in a two dimensional system in the
finite temperature renormalized classical regime.

It is assumed that the N\'eel order parameter fluctuations 
in the charge ordered stripe phase are governed by an AQNLS
model\cite{CH}\cite{Parola},
\begin{eqnarray}
\label{AQNLS}
S_{AQNLS} & = & \frac{1}{2 g_0}\int_0^{u} {\rm d}\tau \int {\rm d}^2x
\left( \alpha\left(\partial_x \hat{n}\right)^2+
 \left(\partial_y \hat{n}\right)^2 \right. \nonumber \\
 & & \left. + \frac{2}{1+\alpha}\left(\partial_{\tau}\hat{n}\right)^2\right).
\end{eqnarray}
where the bare coupling constant $g_0$ and the spin-wave velocity $c$ 
are those of the isotropic system, while $\alpha$ parametrizes the anisotropy.
In the classical limit, this describes spin waves with velocity
$c_y(\alpha)= c \sqrt{(1+\alpha)/2}$ and $c_x(\alpha)=\sqrt{\alpha}c_y(\alpha)$
in the $y$- and $x$ directions, respectively. The slab thickness in the
imaginary time direction $u$ is given by $\beta\hbar c \Lambda$, 
where $\Lambda$ is the cut-off of our spherical Brillouin zone.This model is derived by
taking the naive continuum limit of a Heisenberg model with exchange couplings
$J$ and $\alpha J$ in the $y$- and $x$ directions, respectively. 

The renormalization of this model has received some attention recently
\cite{CH,affleck}. We adopt here a variation on the procedure 
as proposed by Affleck\cite{affleck}. The central observation is that
this model contains two ultraviolet cut-offs. As a ramification of the
anisotropy, the highest momentum states in the x-direction will have an
energy $E^{\rm max}_x$ which is a factor $\sqrt{\alpha}$ smaller than that
of the highest momentum states in the y direction. Therefore, the initial
renormalization flow from $E^{\rm max}_y$ down to $E^{\rm max}_x$ is governed
by one dimensional fluctuations. At  $E^{\rm max}_x$ the resulting model
can be rescaled to become isotropic, albeit with `bare' parameters which are
dressed up by the one dimensional high energy fluctuations. 

Keeping the full model Eq. (\ref{AQNLS}), the
one dimensional fluctuations are integrated out (using momentum-shell
renormalization\cite{CHN})
by neglecting the dispersions
in the x direction entirely. This causes the anisotropy parameter $\alpha$
to become a running variable as well, which is always relevant. When 
the renormalized $\alpha =1$, the model has become isotropic, albeit with
renormalized bare coupling constants.
 
Writing $\hat{n}=(\vec{\pi},\sigma)$, where $\sigma$ is
the component of $\hat{n}$ in the direction of ordering, we expand to one-loop order 
in $\vec{\pi}$. Subsequently, we Fourier transform the $\vec{\pi}$-fields according 
to
\begin{eqnarray}
\vec{\pi}(\vec{x},\tau) & = & \sum_{n=-\infty}^{\infty} \int \frac{{\rm d}^2k}{(2 \pi)^2}
\vec{\pi}(\vec{k},n){\rm e}^{\imath \vec{k}\cdot\vec{x}-\imath\omega_n\tau},
\end{eqnarray}
where $\omega_n=2\pi n/u$ are the Matsubara frequencies. The momenta $k$ are rescaled 
with $\Lambda$ to become dimensionless. Separating the fields according
to 
\begin{equation}
\vec{\pi}(\vec{k},n)=\left\{\begin{array}{lcl} \vec{\pi}_>( \vec{k},n)
 & ; & {\rm e}^{-l}< |k_y|<1 \\
 \vec{\pi}_<( \vec{k},n) & ; & 0<|k_y|<{\rm e}^{-l} \end{array}\right. ,
\label{sep}
\end{equation} 
where $l$ is small, we integrate out the fields $\pi_>$, using a square
Brillouin zone for convenience. Rescaling $k_y$,
$\pi_<$, $u$, $g$ and $\alpha$, we find that the model scales to
larger $\alpha$ (smaller anisotropy). We obtain the following flow equations
\begin{eqnarray}
\alpha  & = & \alpha_0 {\rm e}^{2l},\label{a} \\
\frac{\partial g}{\partial l} & = & -\frac{\alpha}{1+\alpha}g + g^2 I,
\label{dg} \\
\frac{\partial t}{\partial l} & = & t +t g I, 
\label{dt}
\end{eqnarray}
where
\begin{equation}
I = \frac{\sqrt{1+\alpha}}{4\sqrt{2}\pi^2} \int_{-1}^1{\rm d} k_x 
\frac{\coth\left(\frac{u}{2}\sqrt{\frac{1+\alpha}{2}}\sqrt{\alpha k_x^2+1}\right)}{
\sqrt{\alpha k_x^2+1}},
\end{equation}
and where $t$ is the dimensionless temperature, $t_0=k_{\rm B} T/\rho_s^0$.  
From eq. (\ref{dg}) and (\ref{dt}), we find for the slab thickness
$u=g/t$
\begin{equation}
u=u_0\sqrt{\frac{1+\alpha_0}{1+\alpha}}{\rm e}^{-l}.
\end{equation}

From Eq. (\ref{a}) it follows that $\alpha =1$ corresponds with  
$l=l_1=-\ln \sqrt{\alpha_0}$. The bare coupling constant (at $T=0$)
and bare slab thickness
of the effective isotropic model follow by integrating Eq.'s (\ref{dg},\ref{dt})
down to $l_1$ ($g_1 = g(l_1), t_1 = t(l_1)$),
\begin{eqnarray}
g_1 & =  &  g_0/\left[ \sqrt{\frac{2}{1+\alpha_0}}-\frac{g_0}{2 \pi^2}\left(
{\rm arsinh}(\sqrt{\alpha_0})/\sqrt{\alpha_0}+ \right.\right.
\nonumber \\
& & \left.\left.
\ln(1+\sqrt{1+\alpha_0})-
\ln(\sqrt{\alpha_0}(1+\sqrt{\alpha_0})^2) \right) \right].
\label{g1}\\
g_1/t_1 & = & (g_0/t_0)\sqrt{\alpha_0(1+\alpha_0)/2} 
\label{u1}
\end{eqnarray}
Except for these altered bare quantities, the isotropic model is analyzed in 
the standard way\cite{CHN}. 

Putting $g_1=g_c=4\pi$ and solving $g_0$, we find the critical bare
coupling for the anisotropic model
\begin{eqnarray}
g_c(\alpha_0) & = & 4 \pi \sqrt{\frac{2}{1+\alpha_0}}/\left[1+\frac{2}{\pi}\left(
{\rm arsinh}(\sqrt{\alpha_0})/\sqrt{\alpha_0}+ \right.\right. \nonumber \\
 & & \left.\left. \ln(1+\sqrt{1+\alpha_0})-
\ln(\sqrt{\alpha_0}(1+\sqrt{\alpha_0})^2) \right) \right].
\end{eqnarray}
We find this result to be the same within a couple of percents  as
the outcome of large N mean field theory\cite{CH}, while the difference
originates in an inaccuracy in our calculation related to  
the switch from the square (at  $E>E_{\rm max}^x$) to the spherical Brillouin zone of the effectively isotropic model.

\begin{figure}[h]
\centering
\hspace{-.11 \hsize}
\epsfxsize=1.1\hsize
\epsffile{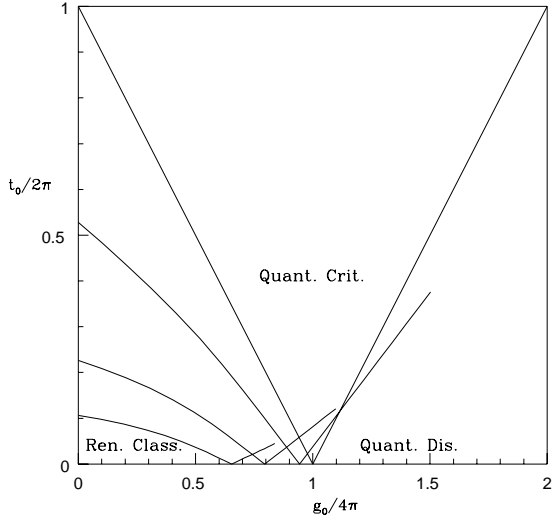}
\caption{Cross-over diagram for the anisotropic QNLS. The lines are for $\alpha=1,0.4,0.1$ and $0.025$ from top to bottom. The end-points of the quantum-critical
to quantum-disordered lines map onto $(g_1,t_1)=(8\pi,2\pi)$.
Notice that when $t_0$ becomes larger than the crossover temperature 
from renormalized classical to quantum critical at $g_0 = 0$ one
dimensional fluctuations are dominating for all values of $g_0$.}  
\label{f1}
\end{figure}

For $\alpha_0=1$, the one-loop cross-over lines
between the QC and the RC/QD
regime are given by $t=\pm 2\pi(1-g/4\pi)$. Taking 
$(g_1,t_1)$ to lie on these lines and iterating the flow equations backwards, we
obtain the cross-over diagram for the anisotropic model, shown in fig.~\ref{f1}.
Note that the
anisotropy has a stronger effect on the $t$-dependence of the RC to QD line
than on its $g$-dependence. This already indicates that the $T=0$ properties
will be less affected by the anisotropy than those at finite temperatures.

The one-loop mapping to an isotropic QNLS provides a simple way of calculating
the correlation length in the anisotropic model. Noting that
the correlation length in the $y$-direction scales as $\xi=\xi_0{\rm e}^{-l}$ under
Eq.~(\ref{sep}), it immediately follows that
$\xi (g_0,t_0)= {\rm e}^{l_1}\xi_{\rm isotr.} (g_1,t_1)$.
Inserting the 1-loop 
expression for $\xi_{\rm isotr.}$ in the RC regime
\cite{CHN} and using Eq.'s (\ref{g1},\ref{u1}) (the use of the $T=0$
expression for $g_1$ Eq.(\ref{g1}) is a good 
approximation if $g_1/t_1\gg 1$), 
\begin{eqnarray}
\xi (g_0,t_0) & = & \frac{0.9}{\sqrt{\alpha_0}} \frac{g_1}{2t_1}\exp\left[\left(1-\frac{g_1}{4\pi}\right)/t_1\right]
\nonumber \\ 
 & \simeq & 0.9 \frac{g_0}{2 t_0} \sqrt{\frac{1+\alpha_0}{2}}\exp\left[\sqrt{\alpha_0} 
\rho_s(0) /k_B T\right].
\label{ksi}
\end{eqnarray}
where the renormalized $T=0$ stiffness is given by,
\begin{equation}
\rho_s(0)=\rho_s^0\left(1-\frac{g_0}{g_c(\alpha_0)}\right).
\label{restiff}
\end{equation}
Eq.'s (\ref{ksi},\ref{restiff}) is our central result. It shows that the correlation length
in the renormalized classical regime has a twofold {\em exponential}
dependence on the anisotropy, both originating in the high frequency one
dimensional fluctuations. As already pointed out by CH\cite{CH}, 
the anisotropy causes
$g_c$ to decrease (e.g., Fig. 1), 
leading to a reduction of $\xi$ at a given temperature.
However, we find an additional $\sqrt{\alpha}$ in the exponent which has
been overlooked before, and this is the specific way in which the 
greater effect of the thermal fluctuations as we noted earlier  shows up
in the renormalized classical regime. In fact, it shows that the basic
invention of CHN\cite{CHN}
 is straightforwardly extended to the anisotropic case.
The correlation length is given by the expression for the classical
system, and quantum mechanics only enters in the form of a redefinition
of the stiffness. However, for the classical correlation length expression
one should use the one for the {\em anisotropic classical model}.
Using the same procedure as  
for the quantum model, it is easy to demonstrate
 that the correlation length 
of the anisotropic classical $O(3)$ model in 2D behaves as $\xi \sim 
\exp ({\sqrt{\alpha_0} \rho^0_s / k_B T})$, and this explains the occurrence 
of the additional $\sqrt{\alpha_0}$ factor.

The finiteness of the N\'eel temperature is caused by  small intraplanar
spin-anisotropies and interplanar couplings. 
Keimer {\em et al} \cite{Keimer} have shown 
that in $La_2CuO_4$ the former dominate, and
these can be lumbed together in a single term $\alpha_{eff}$ which plays
the role of an effective staggered field. 
The N\'eel temperature can be estimated by comparing the thermal energy
 $k_{\rm B}T_N$ to the energy-cost of
flipping all spins in a region the size of the correlation length 
in the presence of the effective staggered field. 
\begin{equation}
k_{\rm B}T_N(\alpha) \simeq J \alpha_{\rm eff}\left(\frac{\xi(T_N,\alpha)}{a}
\frac{M_s}{M_0}\right)^2.
\label{TN}
\end{equation}
Because it is not expected that stripes will influence the spin anisotropies
strongly, we can use the estimate for $\alpha_{\rm eff}$ as determined for the
half-filled system: $\alpha_{\rm eff} = 6.5\times 10^{-4}$ \cite{Keimer}.  
For our estimate of $T_N$,
we will use spin-wave results for the renormalized stiffness, susceptibility and
spin wave velocity \cite{Igar}. For $S=1/2$, they are $\hbar c=0.5897\sqrt{8}J a$, 
$\chi_{\perp}(0)=0.514 \hbar^2/(8 J a^2) $, and  $\rho_s=c^2
\chi_{\perp}(0)$. The bare coupling constant is obtained from $(g_0/4\pi)=1/(1+4\pi\chi_{\perp} c/\hbar\Lambda)$ \cite{CHN}, which yields
$g_0=9.107$ for $\Lambda a=2 \sqrt{\pi}$. We notice that the 1-loop result
for the prefactor is not correct, but this factor is not very 
important as far as the reduction of the N\'eel temperature is concerned.

\begin{figure}[h]
\hspace{-.05 \hsize}
\epsfxsize= \hsize
\epsffile{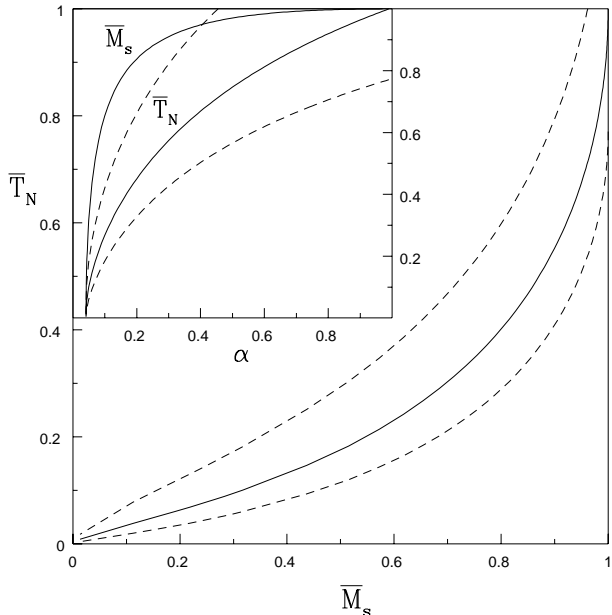}
\caption{The N\'eel temperature versus the zero temperature
 staggered magnetization, with the anisotropy as implicit parameter. Both
quantities are normalized with respect to their value
in the isotropic system. The upper/lower dashed line gives $\bar{T}_N$
for $\alpha_{\rm eff}$ (N\'eel stabilizing field)
a factor 10 larger/smaller than in the isotropic system. {\em Inset}:
$\bar{T}_N$ and $\bar{M}_s$ as a function of the anisotropy parameter
$\alpha$.}
\label{f2}
\end{figure}

Since our $T=0$ results coincide with those obtained by CH\cite{CH}, 
we use their expression for the zero temperature staggered 
magnetization\cite{topology},
\begin{equation}
\frac{M_s(\alpha)}{M_s(1)} = \sqrt{\frac{1-g_0/g_c(\alpha)}{1-g_0/4\pi}},
\label{magn}
\end{equation}
and its anisotropy dependence is shown together with the results for the N\'eel temperature in the inset in fig.~\ref{f2}. To illustrate the effects of a different $\alpha_{\rm eff}$ in the stripe phase
(e.g.,  $J_{\perp}$ may be much reduced due to frustration) 
we have also
plotted the results for $\alpha_{\rm eff}(\alpha<1)= 10 \alpha_{\rm eff}(\alpha=1)$
(upper dashed line) and for  $\alpha_{\rm eff}(\alpha<1)= 0.1 \alpha_{\rm eff}(\alpha=1)$ (lower dashed line). 
In Fig.~\ref{f2} $T_N$ is plotted
versus $M_s$. As expected, the dependence of $T_N$ on anisotropy is considerably
stronger than that of $M_s$. A reduction of $M_s$ by a factor of 2 due to
a spin-wave anisotropy $\sim \sqrt{\alpha} \sim 1/4$ 
order is accompanied by a suppression of $T_N$ by roughly an order of magnitude. 

In the above we relate different experimentally accessible quantities
(spatial- and spin anisotropies, N\'eel temperature, $T=0$ staggered
order, correlation length) and further\cite{tranquada1,buechner}
experimentation is needed to unambiguously demonstrate that spatial
anisotropy is the cause of the low spin ordering temperature. If the
fluctuation behavior in the RC regime is indeed as general
as suggested by the present analysis, other sources of stripe
induced spin disorder could have similar consequences. For instance,
local charge deficiencies in the stripes caused by quenched disorder
would give rise to unscreened (by charge) pieces of domain walls. Such
stripe defects are like the dipolar defects discussed by Aharony {\em et al},
\cite{Ahrony} and their frustrating effect is expected to be disproportionally
stronger at finite temperature than at zero temperature. 

Above all, the present analysis shows that a Landau mean-field analysis falls
short as a description for the thermodynamic behavior of the stripe phase
because of the importance of fluctuations. Stronger, thermodynamics does
not offer an unambiguous guidance regarding the microscopy (frustrated 
phase separation \cite{emkiv} versus `holon' type mechanisms\cite{zagu}).   
Here we have focussed on the transversal spin fluctuations, and given that 
there is ample evidence for a pronounced slowing down of the spin dynamics 
at the charge ordering temperature,
these undoubtedly play an important role. It is
noted that recent results point at a similarly important role of fluctuations
in the charge sector\cite{lee}.

{\em Acknowledgements.} We thank V. J. Emery, B. I. Halperin,
S. A. Kivelson and W. van Saarloos for 
stimulating discussions. Financial support was provided by the Foundation
of Fundamental Research on Matter (FOM), which is sponsored by the
Netherlands Organization of Pure research (NWO). JZ acknowledges support
by the Dutch Academy of Sciences (KNAW).

\end{multicols}


\begin{references}
\bibitem{eq13note} Following CHN\cite{CHN}, Eq.'s (\ref{ksi},\ref{restiff}) 
can also be obtained 
by integrating out all
quantum fluctuations in the renormalized classical region to 1-loop order.
This yields a 2d anisotropic classical non-linear sigma model, with
effective dimensionless temperature 
$$
\frac{1}{t_0} = \frac{1}{k_{\rm B}T}\left[\rho_s(0)+\frac{k_{\rm B}T}{2 \pi\sqrt{\alpha_0}}
\ln\left(\frac{\hbar c \Lambda}{k_{\rm B}T}\sqrt{\frac{1+\alpha_0}{2}}\right)
+{\cal O}(T^2)\right].
$$
where $g_c(\alpha_0)$ is now precisely the critical coupling obtained by CH.
Momentum-shell renormalization of this 2d model again yields eq.(\ref{ksi}).
\bibitem{topology} Because of the neglect of topological terms, this
expression yields a lower bound for the staggered magnetization.
\end{references}
\end{document}